# Combined Light Excitation and Scanning Gate Microscopy on Heterostructure Nanowire Photovoltaic Devices

Yen-Po Liu[1,3], Jonatan Fast[2,3], Yang Chen[2,3], Ren Zhe[1,3], Adam Burke[2,3], Rainer Timm[1,3], Heiner Linke[2,3], Anders Mikkelsen[1,3]

[1]*Division of Synchrotron radiation, Department of Physics, Lund University, Sweden*

[2]*Division of Solid State Physics, Department of Physics, Lund University, Sweden*

[3]*NanoLund, Lund University, Sweden*

## Abstract

Nanoscale optoelectronic components achieve functionality via spatial variation in electronic structure induced by composition, defects, and dopants. To dynamically change the local band alignment and influence defect states, a scanning gate electrode is highly useful. However, this technique is rarely combined with photoexcitation by a controlled external light source. We explore a setup that combines several types of light excitation with high resolution scanning gate and atomic force microscopy (SGM/AFM). We apply the technique to InAs nanowires with an atomic scale defined InP segment, that have attracted considerable attention for studies of hot carrier devices. Using AFM we image the topography of the nanowire device. SGM measurements without light excitation show how current profiles can be influenced by local gating near the InP segment. Modelling of the tip and nanowire can well predict the results based on the axial band structure variation and an asymmetric tip. SGM studies including light excitation are then performed using both a white light LED and laser diodes at 515 and 780nm. Both negative and positive photoconductance can be observed and the combined effect of light excitation and local gating is observed. SGM can then be used to discriminate between effects related to the wire axial compositional structure and surface states. The setup explored in the current work has significant advantages to study optoelectronics at realistic conditions and with rapid turnover.

# Introduction

Spatial variation in band gap, band alignment and electrical/optical active defects in low dimensional device structures are central for their function. To understand and even operate such devices spatiotemporal variation of the electronic structure is important which can be introduced by electrostatically gating the devices. To dynamically probe these properties in nanostructures, the use of a local moveable gate is an excellent tool. Scanning gate microscopy (SGM), which measures changes in the conductance of the device as a function of the local gating effect produced by an STM or AFM tip close to the nanostructure, provides a link between performance and local electronic structure changes. SGM is especially useful to study nanowire (NW) devices as the (capacitive) coupling between tip and sample reveal both surface and bulk properties of a wire. This provides information such as the local carrier type and a qualitative indication of the carrier concentration in the NW under the tip. SGM has been used to study semiconductor NWs at low temperature[1–3], to study carbon nanotubes[4–8] and other low dimensional systems[9–16]. However, while SGM has been used to study transport and low temperature quantum effects, it has rarely been used to study optoelectronic components. To apply SGM on optoelectronic devices, the combination with light excitation is relevant to directly study how photoexcitation and subsequent carrier transport and relaxation is affected by the spatial variation of the electronic structure. In most previous works combining scanning probes and optical sources a focus has been on using the tip structure to induce local field enhancement of the light in NWs[17–19]. While this can give information on the effect of light excitation in specific parts of a structure, using SGM provides a complimentary understanding on how changing the material electronic structure locally affect the light absorption and subsequent carrier transport and relaxation properties. Thus, the SGM data can potentially be used to rationally design the composition and geometry of the device for better performance.

Semiconductor NWs are promising candidates for a wide range of optoelectronic applications including solar cells[20–25], photodetectors[26,27], LEDs[28–30] and nanophotonic neuromorphic computing[31]. Recently, the extraction of photoexcited carriers with energy above the conduction band minimum, so called hot carrier devices, has been explored[32]. This promising way to increase energy output of a solarcell can be well investigated in NWs as the 1D transport along the wire and the opportunity for axial band gap engineering is excellent for developing such device concepts. As a result, InAs NWs with/without InP segments, as used in the current work, has been explored in a number of hot electron studies [17,32–34].

In the present work, nanometer precise SGM is combined with light excitation using both broad band white LEDs as well as single-mode lasers with a sharp photon energy bandwidth, to measure heterojunction NW devices. We study how the current-voltage curves of an InAs NW with an InP segment is influenced by local gating effects, and photon excitation. This combination is relevant for understanding the behavior of novel opto-electronics. With its larger band the InP segment acts as a barrier for electron and hole transport along the wire, which is then strongly altered by the SGM as it is placed on top of the segment either raising or lowering it. While light excitation can lead to additional carriers in the wire increasing the observed currents, the light can also excite defect states which effectively create a negative photoconductivity. Both instances are observed, depending on surface chemistry, and can be investigated in the SGM setup. Using different light sources in combination with the SGM studies enables the probing of both broadband excitation as well as excitation at specific wavelengths. This is the first work reporting SGM in combination with laser/LED light sources on such an InAs/InP energy barrier NW device.

## Material and Methods

### Experimental setup

SGM and AFM measurements were carried out using an Scanning probe microscopy (SPM) system (from Unisoku Co., Ltd.) designed for combined scanning probe imaging and easy optical signal excitation/extraction. The basic setup is shown in Figure 1(a). It provides tip-movement precision to the sub-nanometer scale and allows four-channel electrical measurement on the sample holder for simultaneous device electrical measurements. To have a mobile gate scanning over the InAs/InP NW device, the SPM tip must be conductive, and a q-plus AFM tip is used for the setup, which runs in air and at room temperature (environmental conditions relevant for photovoltaic applications). For SGM measurements a conductive tip can be biased while still operating in AFM mode. As a result, it can acquire nm-scale topography while the device current, influenced by the biased-tip gate, is recorded simultaneously.

The implementation of the laser/LED system is not trivial because the AFM tip must be positioned on top of the device, and would thus preclude the laser beam to enter with a normal incidence angle as was done previously in OBIC measurements using the same setup[32]. Instead, the optics is tilted 30 degrees so that the focused beam comes with an angle to avoid blocking, as shown in the illustration of the setup is shown in Figure 1 (a). As a result, the beam can still be focused to the diffraction limit[32] in one direction, but will be significantly defocused in the other direction. However, even with tilted optics it is possible to observe the tip and the device as seen in Fig S2 of the SI. Single-mode fiber lasers, Thorlabs LP515-SF3 and LP785-SF20, were guided into a collimator lens F810FC-780 and then pass through two two-way mirrors to couple the optical lightening LED, Thorlabs MNWHL4, and the camera for illustrating and aligning. The optical setup illustration is shown in Fig S1. The laser is focused with a 20x objective, Olympus LMPLFLN 20x, to have enough space for the AFM tip sitting in between the objective and the device. The full optics schematic graph is shown in the SI. For better focusing a long working distance 100 x objective can be inserted to reach the diffraction limit, however in the present case we only used the 20x objective, which results in a broad laser beam with a spot size significantly larger than the device making alignment of the light beam fast. While the system can be used to achieved diffraction limited spot size in one direction with a broad spot in the other using the 100 x lens asymmetric illumination this option was not used in the present work. The setup and optics schematics can be seen in the Supplementary Information (SI).

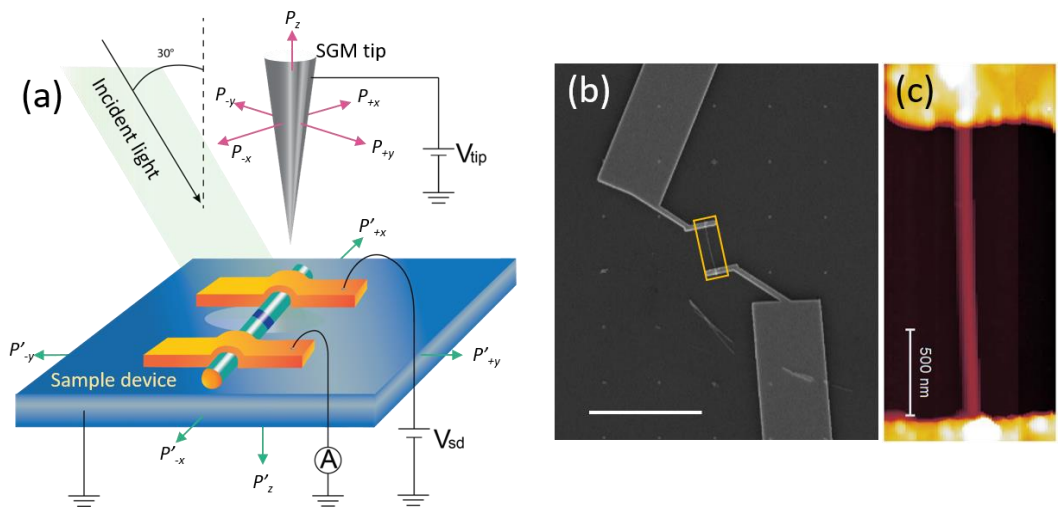

*Figure 1. Illustration of the combined SGM and ligth excitation setup. (a) schematic of the InAs NW with InP segment device scanned under a mobile gate tip $V_{tip}$, homogeneous laser and external bias $V_{sd}$ Both tip and sample can be scanned as indicated with the P marked arrows. (b) SEM image of the InAs/InP barrier NW device with a white scale bar of 5 μm and a yellow rectangular showing where the AFM image in (c) was taken (c) AFM image of the device taken simultaneously with the SGM measurement.*

**Device fabrication**

The InAs NWs with a 25 nm InP segment have a total length of 2 µm and a diameter of about 50 nm. The resulting potential barrier of 0.56/0.38 eV[35,36] in the conduction/valence band respectively is shown in Figure 2(d). The NWs are grown by chemical beam epitaxy using gold seed particles deposited on an InAs (111)B surface, and then mechanically deposited from growth substrate to the sample substrate. InAs and InP segments are high quality crystalline wurtzite structure and weakly n-type[37,38]. After transferring the NWs to a n-type Si substrate, with a 100 nm SiO2 layer on top. Electron beam lithography was used to make contact electrodes, 25 nm Ni and 75 nm Au. Surface sulfur passivation[39] well known to produce InAs nanowire-metal contacts of Ohmic quality[40]. A SEM image of the InAs/InP device can be seen in Figure 1(b).

**Device Measurement**

First, we navigated the AFM to the device area using the optical microscope and imaged the whole device within the scanning range of the AFM. The device current was then measured simultaneously with AFM imaging to acquire the device current as well as topography images. Compared to standard SPM on few nm flat samples, AFM scanning across the full device with 100 nm high electrodes may cause the instability of the qPlus tip, therefore, slow scanning with high gain factors was necessary. Two different types of qPlus tips were used. For initial measurements broader tips were used, but later we used sharper tips that could provide a more local gating effect, as shown in Figure S5 (a) and (b), respectively. Since the biased AFM tip is positioned extremely close to the sample, it provides high electrical fields near the surface making the AFM scanning less stable when scanning a larger range covering the whole device. To compensate this, we scanned only the NW region and not a major part of the electrodes while using the sharper tip. The LED illumination was performed with a Photonic Optics Hi power spot 5800K LED, which spectrum is shown in Figure S6. The laser illumination conditions are at 0, 1 and 2 mW using 515 nm and 780 nm single-mode lasers. The intensity of the laser power was measured with an intensity meter after the fiber and before the collimator, and the transmission rate through all the optical parts of the two wavelength is assumed to be the same because all the optical parts are compatible for 400-800 nm and show constant rate within the wavelength range.

## Results and discussions

Initially SGM measurements were performed on a device in a dark environment to have a reference for further measurements and to understand the basic gating influence on the InAs/InP NW device. In Figure 2(a-c), an example of a SGM measurement is shown illustrating what can be achieved. From Figure 2(a) we clearly see the NW between its contacts demonstrating the high topographical resolution of the AFM. Figure 2(b) shows a very strong, highly localized response of the current when the gated tip is above the InP barrier. From this image the axial local gating profile exactly along and on top of the NW can be obtained as seen in Figure 2(c). Going into more detail, the map of the electrical current along the NW as a function of the tip position, will generally be broadened by the width of electrostatic potential created by the tip, which will depend on the tip shape. The AFM image is a convolution of the tip and the NW device, and in this case a double tip effect is seen as a shadow next to the main NW structure (NW device structure is known from SEM images). Because the double tip is perpendicular to the wire axis it will mostly influence the gate induced current profile across the NW. However, the finite tip shape will generally broaden the measured SGM profiles and the potential applied by the tip falls of rather slowly. As a result, one would usually expect a current variation which is broader than the InP segment as is seen in Figure 2(b). Additionally, as the tip can influence molecules from the air adsorbed on the NW or defect states in the NW native oxide, which can affect conductivity[41–45], this can result in some additional noise in the measurements. This can be seen as a variation in the current signal for each scan and is difficult to avoid in an ambient air setup. However, the overall signal features are reproducible. For each SGM scan an image was recorded when the tip was moving from left to right (forward) and one from right to left (backward). Comparing profiles from such images (see Figure S7 in SI) we can observe that the SGM profiles are qualitatively reproducible, although some noise is observed presumably due to changes in the surface region as noted before. To plot the average current profile along the NW we identify the corresponding topographic area of the NW in the AFM image (blue box in Figure 2a). Using this a current profile along the device (averaging 12 image lines in this case) can be extracted from the SGM image as is shown in Figure 2(c). This is then the current profile when the tip is on top of the wire, which should be most easily comparable with theoretical modelling. In Figure 2, (with a positive tip voltage) a clear current change is observed along the wire consisting of an asymmetric peak followed by a small dip along the NW. A graphic illustration of the band diagram of the NW with $V_{SD}$ is shown in Figure 2(d). The large current fluctuation is assumed to be at where the InP segment is located. A direct visual location of the InP segment in a specific device is not possible as the diameter difference of the InP segment compared to the surrounding InAs is very small. However, as the position of the InP segment (in the middle of the NWs) is known from previous work on the same NWs (and has very good reproducibility) we find that the observed position agrees with what would be expected from the known geometry of the wire. Thus, after identifying the InP segment with the SGM tip we can go on and probe the wire in more detail as will be done below. Before reaching this point, we explore the resolution a bit further. In Figure 2(e) we show results with another tip scanned across the NW instead of along the NW and the side-facets of the NW (the wire is hexagonal) can now be seen. This is a natural consequence of the high resolution of this type of SGM system as discussed previously[1]. Again this emphasize that we get responses both when the tip is on top and on the side of the NW, so to have a consistent measurement we only use the signal when the tip is on top of the NW.

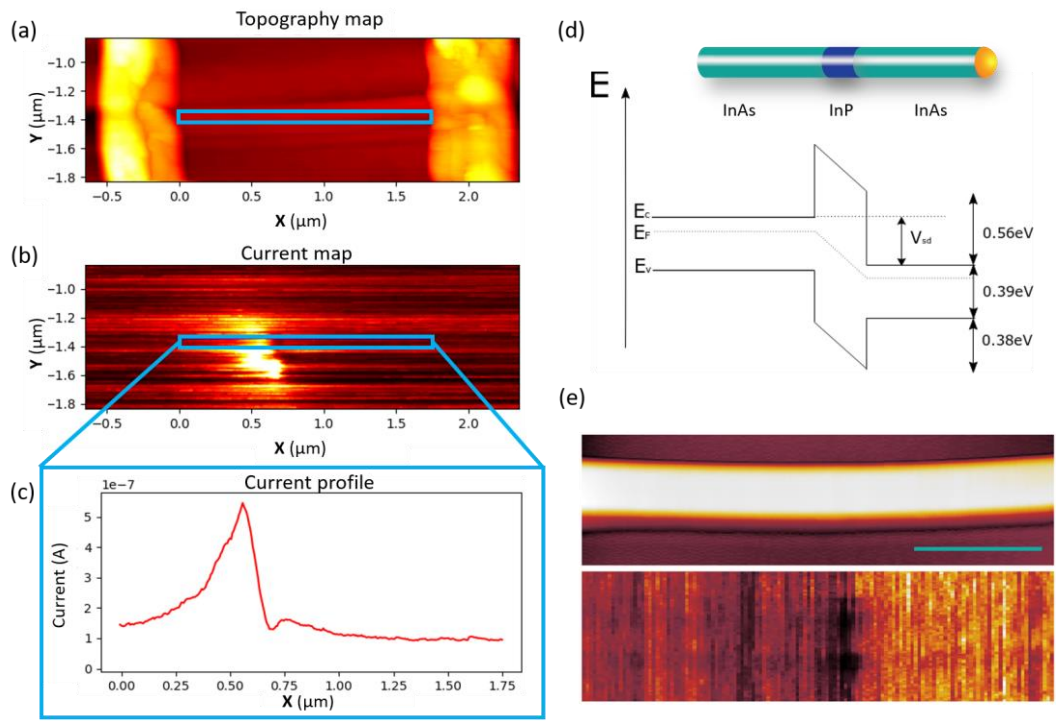

*Figure 2. SGM data measured with $V_{SD}$=+0.6V and $V_{tip}$=+1.0V in dark. (a) Topography map measured with the q-plus AFM (b) Device current map measured simultaneously with topography (c) Device current profile, averaging 12 lines, corresponding to the blue squares, where the NW is located. (d) Band diagram of the InAs/InP NW with a $V_{SD}$ applied. (e) Another set of SGM on device 2, the same type of device, with focus scanning on the NW excluding scan over the electrodes for higher stabiity. Upper: AFM image, lower: device current map. The green scale bar is 500 nm.*

Figure 3 (a) show the current profiles extracted from SGM images under different tip biases giving an overview of the behavior at different local electric fields. The source-drain bias $V_{SD}$ is kept at +0.6 V, the tip bias is varied from -1 V to +1.5 V with step sizes of 0.5 V. The current profile with the highest peak is measured with a tip bias of +1.5 V and here it can be noted that the peak shape is not symmetric. The asymmetric behavior is seen on every curve from negative to positive tip bias, including the curve in Figure 2 (c), and the dip after the peak is found in all positive tip bias curves. For negative tip biases a broad valley in the current is seen instead of the peak at positive tip biases. The variation in the current increases again on the InAs segments at the tip bias -1 V. As the variation is consistent in forward and backward images it is not due to desorption, but likely an indication of de-trapping process with the higher negative tip bias gating. The behavior at higher voltage is however attributed to the InP barrier as will be discussed now. Measurements were also performed with -0.6V bias on the NW with similar results for the behavior observed across the barrier region (considering the lower voltage between the tip and the wire). This would indicate that the behavior seen at the barrier is related to the fundamental symmetric geometry of the NW.

To understand the general behavior, we have modelled the influence of the tip on the NW system. Comsol multiphysics 2D simulations were performed on a 1.4 µm long InAs NW with a 25 nm InP segment with known values for bandgap, band alignment and other relevant material parameters in a partial differential equation (PDE) model [46]. The tip was included as floating gate placed at different positions outside the NW and with a specific shape to be defined and its electrostatic influence on the wire would be included. The contacts at the end of the NW were assumed to be similar and ohmic and possible surface states are not included in the simulations. The equation for the current density is:

$$\boldsymbol{J_n = q\mu_n n E + q D_n \nabla n}$$

The device current is calculated integrating the carrier density along the wire. Calculations were carried out under several different conditions varying tip shape or tip-sample distance as these are not exactly known. Using a triangular tip with a width at the top of 300 nm on top and length of 100 nm ending at the NW (see inset of Figure 3b). The width of the tip at the bottom near the wire is 100 nm and the distance to sample is 10 nm. With these (physically reasonable) parameters a good agreement with the experimental results could be achieved as seen in Figure 3(b). Both the peak shape and the tip voltage dependent behavior could be reproduced well. In particular, the asymmetric shape of the tip was important to observe the shape of the profile, while the qualitative voltage dependent behavior was observed over a wide range of parameters and with a symmetric tip. As obtaining a perfectly symmetric tip is not always possible and tip condition will be varying over time, it is a relevant observation that while the tip shape influence the observed current profiles, this can be included in the simulation and thus well understood in the analysis. While not necessary in the present case (as the geometry of the barrier is already know), the tip shape can also be inferred independently using crystalline structures with a well know shape or SEM.[47] The modeling results indicates that the major features observed in the SGM data in Figure 3(a) are derived from the electrostatic interaction of the tip with the NW, lowering or raising the InP barrier with respect to the InAs conductors on each side which leads to a larger or smaller limitation of the current flow across the NW. The reason for the small dip at positive tip voltages seen after the barrier cannot be well explained by the model, although some hint of a decrease is seen. However, the conductivity of the NW is strongly influence by surface states which can also significantly influence the conductivity across the InP barrier as has previously been observed.[48] As the tip gating is known to influence such states this can lead to a dip in the current.

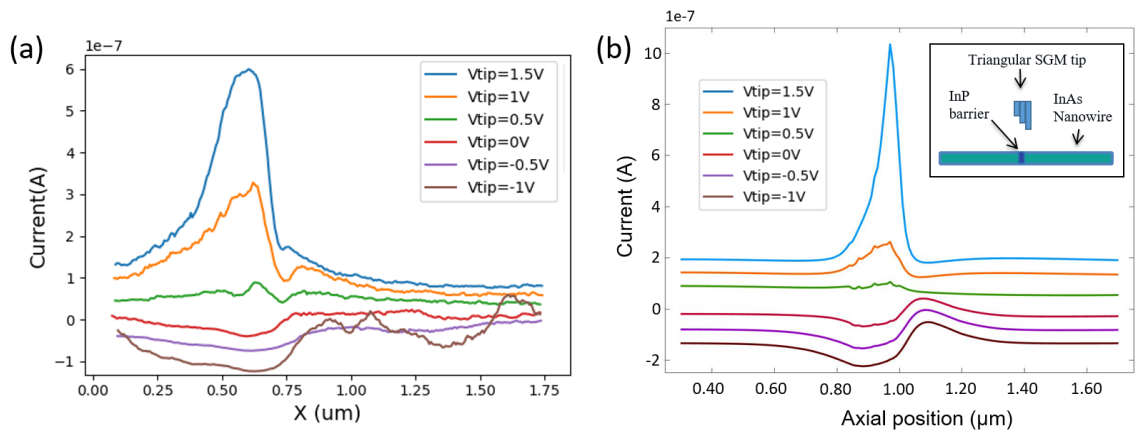

*Figure 3.* Experimental and simulated SGM data on the device (a) water fall plot for SGM data of $V_{SD}$=+0.6V and $V_{tip}$ from +1.5 V to -1 V with step size 0.5 V. The current scale is aligned for $V_{tip}$=+1.5 V curve, and small offsets of 50, 50, 80, 110, 140 nA were applied from $V_{tip}$=+1 V to -1 V, individually. (b) Computational SGM simulation of the InAs/InP NW device with $V_{SD}$=+0.6V from tip bias -1 V to +1.5 V, 0.5 V per step. Offsets of 50 nA are put between each line from +1.5 V to -1 V. The tip in the simulation is asymmetric. Inset: the graphic of the simulation model with the scales described in the text. The shape of the tip in the figure is for illustration purpose and not real scale.

The current profile of experimental data with tip bias -1 V shows a significantly varying current profile even beyond the InP segment and the current around the contact increased much. We ran the measurement several times, and the phenomenon is reproducible. This behavior cannot be explained by our model. However, the reason for this is likely that the electrons at surface states are de-trapped stimulated by gating[49,50]. This will become further relevant when we move to photon excitation. During the SGM measurement with negative tip bias, the trapped electrons can be released back to the valence band by an appropriate negative tip bias, for example, -1 V, scanning over to de-trap them by repulsion. The effect of de-trapping process gaining higher electron current is more effective when the negative biased tip is next to the contact electrodes, as shown in Figure S7. This effect can also be seen when using the substrate as a backgate as shown in Figure S4 in SI.

After analyzing the NWs under dark conditions, we move to SGM imaging of the same NW while exciting it by a standard white LED light source with a spectrum as seen in Fig S6. First, in Figure 4a we show the I(V) curve of this NW device. As can be seen, in the present case the light excitation leads to a diminishing of the observed current corresponding to a negative photoconductivity (NPC) behavior. To further understand this behavior, we performed SGM measurements of the NW as before and show results for +1V, 0V, -0.5V in Figure 4(b) using the same tip, as shown in Figure 5(a). While the current is lower with the illumination, the qualitative behavior is in fact similar to what was observed with no light in the region around the barrier. The only change in that region is that the dip after the large peak observed at +1V on the tip is not observed when the light is on. This would indicate that the change in I(V) behavior observed in Figure 4(a) is likely not directly related to the barrier as we would then expect a qualitative change in the SGM observations. Instead, we propose that the change is due to changes in surface trap states due to the light. As found previously[32], illumination of the device produces hot carriers in the InAs and for asymmetric illumination this can be observed as a current in the device[17,32]. However, qualitative different photoconductance behavior can likely be found for the InAs/InP devices depending on light and surface conditions[48]. The small bandgap III-V semiconductors usually display positive photo-conductance (PPC) behavior; however, due to the fact that the InAs NW has trapping states[49,51,52] above the conduction band, the free electrons excited to the higher states may fall into the trapping state. Hence, lower carrier density in the NW and lower conductance (NPC) can occur in the InAs. The behavior is also intensity dependent and surface chemistry dependent[53,54], for the wire investigated in Figure 3 and Figure 4 we are in this NPC regime. The disappearance of the small dip found at

+1V on the tip would also be consistent with this being surface states that the tip under dark conditions can influence, but which are altered by the light source. At -0.5 V tip voltage, a reduction in current does not occur at the contact regions, which would again be indicative of an influence of the tip on surface states which now changes with light excited trap states. An important observation thus is that using the SGM in combination with light we can begin to entangle the different contribution of the InP barrier and the surface states which cannot be deduced from the pure device I(V).

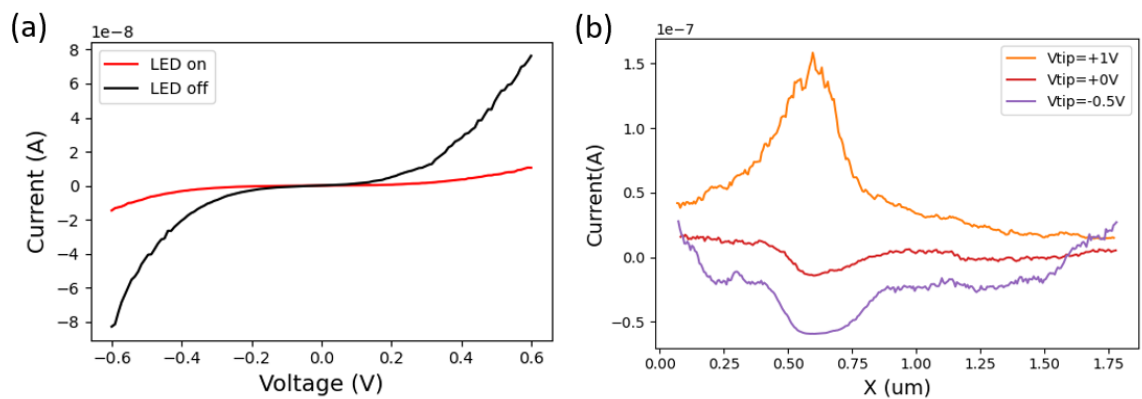

*Figure 4*. SGM with LED illumination (a) schematic of the InAs nanowire with InP segment device scanned under SGM with an LED lamp. (b) SGM current profile with $V_{SD}=+0.6V$ and $V_{tip}=+1.0V$ (orange), 0V (red), and -0.5V (purple) under the LED lamp illumination. There are current offsets to the curves of 0 nA, -20 nA, and -60 nA, respectively.

The spectrum of the LED lamp used for excitation ranges from 400 nm to 750 nm, which is relevant for observing broadband effects found under more realistic conditions. However, it is interesting to also explore separate wavelengths and a wider range of device/tip conditions. To investigate it we carried out measurements with a new set of devices and a new sharper and more symmetric tip. Laser diodes with wavelengths of 515 nm and 780 nm were used for excitation with two different, and known wavelengths. The device and tip used, are shown in Figure S5 (d) and (b), respectively. The laser intensities are measured before the single-mode fiber at the controller unit (the optical component setup is shown in Figure S1), and the intensity loss for 515 nm and 780 nm diodes are assumed to be the same as all the optical components have flat transmission rate curve in the wavelength range of 400 nm to 800 nm.

The overall device response to light is seen in Figure 5 (a), in contrast to the previous device, we observe a positive photoconductance. This must be related to the difference in surface conditions of the NWs which strongly affect conductivity and optical response[53]. While the internal InP barrier remains the same, molecular adsorption on the surface can, for example, change conductivity by orders of magnitude. Additionally, as discussed above, both NPC and PPC behavior has been observed on InAs depending on surface chemistry and light intensities. The strong variation in charge states at the surface (of NW and tip) change the effect of applying the tip potential. The SGM profiles along the NW with illumination of 780 nm and 515 nm laser wavelengths are seen in Figure 5 (b) and (c), and their transport behaviors are quite different. The 515 nm laser, corresponding to 2.4 eV photons, combined with the SGM tip increase the current transported through the NW broadly at the region around the InP segment at +1.5 tip voltage and it has a sharp peak observed at all tip voltages at the InP segment. For excitation with the 780 nm laser, corresponding to 1.6 eV photons, a dip is observed at the InP barrier, which is also observed with no light. An interpretation of the behavior at 515 nm is that the broad peak around the InP segment (extending more than a micrometer at 1.5V tip voltage) is related to gate induced alteration of defect states in the InAs that influence the relaxation of hot carriers. The increase is only seen using the 515 nm laser (and not the lower energy 780 nm) and while it is centered on the InP segment it is much broader. The much sharper peak also observed at the InP segment, could be due to the sharp tip electrically affecting the band alignment at the InP or surface states on the InP. However, as there is little influence by the tip voltage and it is quite constant it is more likely a plasmonic scattering effect of the tip. The dip observed while illuminating with the 780 nm laser light is observed independent of voltage and in the dark. This indicates that the behavior is governed in this case by local charge interaction between the tip and surface, that must partially screen out the effect of the tip bias band alignment of the InP-InAs seen for the first wires. Such effects can indeed occur when the surface of tip and/or NW has significant defect populations. The different tip shape would have significant impact on the transport behavior, especially the sharpness of the tip apex for gating effect. The tip shape comparison is shown in Figure S5 (a) and (b). A sharper tip, which is optimized for low-temperature UHV environment, in (b) gives a higher local electric field with the same bias, which could induce a different tip-sample distance as there will be different repulsing forces by the electrical field. We note that a current increase along the wire is observed which is independent of the InP segment and thus likely again must be due to the InAs segments, where the trapping effect is wavelength dependent.[52]

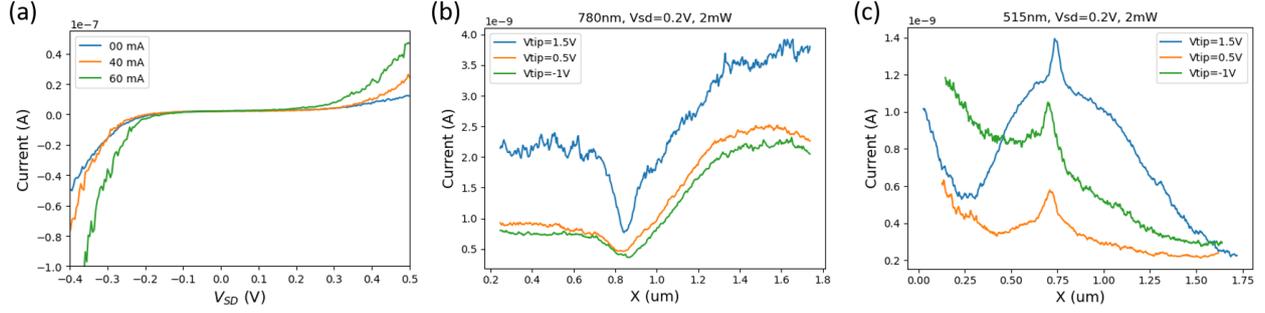

***Figure 5.*** *(a) IV curves of the InAs/InP NW device under 780 nm laser at different light intensities, which are labeled with the controller output current in mA. (b) Laser light combined with SGM with $V_{SD}$=+0.2V and $V_{tip}$ =+1.5 V, 0.5V, -1.0V, 780 nm (1.6ev) illumination at 2 mW. (c) Laser light combined with SGM with $V_{SD}$=+0.2V and $V_{tip}$ =+1.5 V, 0.5V, -1.0V, 515 nm (2.4ev) illumination at 2 mW. The laser intensity is measured after the laser fiber and before the collimator.*

## Conclusions

We have investigated the combination of SGM and light excitation in an ambient air setup that is optimized for measurement under varying excitation conditions and allows fast sample throughput. The moveable gate allows for separation of different effects in semiconductor heterostructures: In particular, we can investigate the different influence of surface states and the effect of the internal compositional heterostructure (InP segment) along the device. We previously found that when illuminating the NW in specific areas along the NW[32], the results can be understood based on the band diagram changes (due to the InP barrier) along the wire. However, the surface trap states on the NW must also be taken into account to explain, for example, the behavior such as both PPC and NPC and all the gating effects. While not explored in the present setup optical beam induced current (OBIC) measurements[32] with a focused light source scanning along the wire could be combined with SGM for further exploration of both surface and lateral geometric effects. While we investigated one type of NW heterostructure the technique is relevant for a wide range of heterostructures as the combined structure at the wire surface with the axial compositional heterostructures along the wire will determine the wire conditions.

# References

(1) Webb, J. L.; Persson, O.; Dick, K. A.; Thelander, C.; Timm, R.; Mikkelsen, A. High Resolution Scanning Gate Microscopy Measurements on InAs/GaSb Nanowire Esaki Diode Devices. *Nano Res.* **2014**, *7* (6), 877–887. https://doi.org/10.1007/s12274-014-0449-4.

(2) Boyd, E. E.; Storm, K.; Samuelson, L.; Westervelt, R. M. Scanning Gate Imaging of Quantum Dots in 1D Ultra-Thin InAs/InP Nanowires. *Nanotechnology* **2011**, *22* (18), 185201. https://doi.org/10.1088/0957-4484/22/18/185201.

(3) Zhukov, A. A.; Volk, Ch.; Winden, A.; Hardtdegen, H.; Schäpers, Th. Distortions of the Coulomb Blockade Conductance Line in Scanning Gate Measurements of Inas Nanowire Based Quantum Dots. *J. Exp. Theor. Phys.* **2013**, *116* (1), 138–144. https://doi.org/10.1134/S1063776112130195.

(4) Hunt, S. R.; Wan, D.; Khalap, V. R.; Corso, B. L.; Collins, P. G. Scanning Gate Spectroscopy and Its Application to Carbon Nanotube Defects. *Nano Lett.* **2011**, *11* (3), 1055–1060. https://doi.org/10.1021/nl103935r.

(5) Freitag, M.; Johnson, A. T.; Kalinin, S. V.; Bonnell, D. A. Role of Single Defects in Electronic Transport through Carbon Nanotube Field-Effect Transistors. *Phys. Rev. Lett.* **2002**, *89* (21), 216801. https://doi.org/10.1103/PhysRevLett.89.216801.

(6) Staii, C.; Johnson, A. T.; Shao, R.; Bonnell, D. A. High Frequency Scanning Gate Microscopy and Local Memory Effect of Carbon Nanotube Transistors. *Nano Lett.* **2005**, *5* (5), 893–896. https://doi.org/10.1021/nl050316a.

(7) Bachtold, A.; Fuhrer, M. S.; Plyasunov, S.; Forero, M.; Anderson, E. H.; Zettl, A.; McEuen, P. L. Scanned Probe Microscopy of Electronic Transport in Carbon Nanotubes. *Phys. Rev. Lett.* **2000**, *84* (26), 6082–6085. https://doi.org/10.1103/PhysRevLett.84.6082.

(8) Kim, Y.; Oh, Y. M.; Park, J.-Y.; Kahng, S.-J. Mapping Potential Landscapes of Semiconducting Carbon Nanotubes with Scanning Gate Microscopy. **2007**, 5.

(9) Aoki, N.; da Cunha, C. R.; Akis, R.; Ferry, D. K.; Ochiai, Y. Imaging of Integer Quantum Hall Edge State in a Quantum Point Contact via Scanning Gate Microscopy. *Phys. Rev. B* **2005**, *72* (15), 155327. https://doi.org/10.1103/PhysRevB.72.155327.

(10) Brun, B.; Martins, F.; Faniel, S.; Hackens, B.; Bachelier, G.; Cavanna, A.; Ulysse, C.; Ouerghi, A.; Gennser, U.; Mailly, D.; Huant, S.; Bayot, V.; Sanquer, M.; Sellier, H.

Wigner and Kondo Physics in Quantum Point Contacts Revealed by Scanning Gate Microscopy. *Nat Commun* **2014**, *5* (1), 4290. https://doi.org/10.1038/ncomms5290.

(11) Connolly, M. R.; Chiou, K. L.; Smith, C. G.; Anderson, D.; Jones, G. A. C.; Lombardo, A.; Fasoli, A.; Ferrari, A. C. Scanning Gate Microscopy of Current-Annealed Single Layer Graphene. *Appl. Phys. Lett.* **2010**, *96* (11), 113501. https://doi.org/10.1063/1.3327829.

(12) Pala, M. G.; Hackens, B.; Martins, F.; Sellier, H.; Bayot, V.; Huant, S.; Ouisse, T. Local Density of States in Mesoscopic Samples from Scanning Gate Microscopy. *Phys. Rev. B* **2008**, *77* (12), 125310. https://doi.org/10.1103/PhysRevB.77.125310.

(13) Jalilian, R.; Jauregui, L. A.; Lopez, G.; Tian, J.; Roecker, C.; Yazdanpanah, M. M.; Cohn, R. W.; Jovanovic, I.; Chen, Y. P. Scanning Gate Microscopy on Graphene: Charge Inhomogeneity and Extrinsic Doping. *Nanotechnology* **2011**, *22* (29), 295705. https://doi.org/10.1088/0957-4484/22/29/295705.

(14) Liu, J.; Cai, Z.; Koley, G. Charge Transport and Trapping in InN Nanowires Investigated by Scanning Probe Microscopy. *Journal of Applied Physics* **2009**, *106* (12), 124907. https://doi.org/10.1063/1.3273380.

(15) Martin, D.; Heinzig, A.; Grube, M.; Geelhaar, L.; Mikolajick, T.; Riechert, H.; Weber, W. M. Direct Probing of Schottky Barriers in Si Nanowire Schottky Barrier Field Effect Transistors. *Phys. Rev. Lett.* **2011**, *107* (21), 216807. https://doi.org/10.1103/PhysRevLett.107.216807.

(16) Aoki, N.; Burke, A.; Cunha, C. R. da; Akis, R.; Ferry, D. K.; Ochiai, Y. Study of Quantum Point Contact via Low Temperature Scanning Gate Microscopy. *J. Phys.: Conf. Ser.* **2006**, *38*, 79–82. https://doi.org/10.1088/1742-6596/38/1/020.

(17) Chen, I.-J.; Limpert, S.; Metaferia, W.; Thelander, C.; Samuelson, L.; Capasso, F.; Burke, A. M.; Linke, H. Hot-Carrier Extraction in Nanowire-Nanoantenna Photovoltaic Devices. *Nano Lett.* **2020**, *20* (6), 4064–4072. https://doi.org/10.1021/acs.nanolett.9b04873.

(18) Arcangeli, A.; Rossella, F.; Tomadin, A.; Xu, J.; Ercolani, D.; Sorba, L.; Beltram, F.; Tredicucci, A.; Polini, M.; Roddaro, S. Gate-Tunable Spatial Modulation of Localized Plasmon Resonances. *Nano Lett.* **2016**, 6.

(19) Zhou, Y.; Chen, R.; Wang, J.; Huang, Y.; Li, M.; Xing, Y.; Duan, J.; Chen, J.; Farrell, J. D.; Xu, H. Q.; Chen, J. Tunable Low Loss 1D Surface Plasmons in InAs Nanowires. *Adv. Mater.* **2018**, *30* (35), 1802551. https://doi.org/10.1002/adma.201802551.

(20) Barrigón, E.; Zhang, Y.; Hrachowina, L.; Otnes, G.; Borgström, M. T. Unravelling Processing Issues of Nanowire-Based Solar Cell Arrays by Use of Electron Beam Induced Current Measurements. *Nano Energy* **2020**, *71*, 104575. https://doi.org/10.1016/j.nanoen.2020.104575.

(21) Hu, Y.; Li, M.; He, J.-J.; LaPierre, R. R. Current Matching and Efficiency Optimization in a Two-Junction Nanowire-on-Silicon Solar Cell. *Nanotechnology* **2013**, *24* (6), 065402. https://doi.org/10.1088/0957-4484/24/6/065402.

(22) Jung, J.-Y.; Guo, Z.; Jee, S.-W.; Um, H.-D.; Park, K.-T.; Lee, J.-H. A Strong Antireflective Solar Cell Prepared by Tapering Silicon Nanowires. *Opt. Express* **2010**, *18* (S3), A286. https://doi.org/10.1364/OE.18.00A286.

(23) Krogstrup, P.; Jørgensen, H. I.; Heiss, M.; Demichel, O.; Holm, J. V.; Aagesen, M.; Nygard, J.; Fontcuberta i Morral, A. Single-Nanowire Solar Cells beyond the Shockley–Queisser Limit. *Nature Photon* **2013**, *7* (4), 306–310. https://doi.org/10.1038/nphoton.2013.32.

(24) Li, H.; Jia, R.; Chen, C.; Xing, Z.; Ding, W.; Meng, Y.; Wu, D.; Liu, X.; Ye, T. Influence of Nanowires Length on Performance of Crystalline Silicon Solar Cell. *Appl. Phys. Lett.* **2011**, *98* (15), 151116. https://doi.org/10.1063/1.3574904.

(25) Zhong, Z.; Li, Z.; Gao, Q.; Li, Z.; Peng, K.; Li, L.; Mokkapati, S.; Vora, K.; Wu, J.; Zhang, G.; Wang, Z.; Fu, L.; Tan, H. H.; Jagadish, C. Efficiency Enhancement of Axial Junction InP Single Nanowire Solar Cells by Dielectric Coating. *Nano Energy* **2016**, *28*, 106–114. https://doi.org/10.1016/j.nanoen.2016.08.032.

(26) Chen, X.; Wang, D.; Wang, T.; Yang, Z.; Zou, X.; Wang, P.; Luo, W.; Li, Q.; Liao, L.; Hu, W.; Wei, Z. Enhanced Photoresponsivity of a GaAs Nanowire Metal-Semiconductor-Metal Photodetector by Adjusting the Fermi Level. *ACS Appl. Mater. Interfaces* **2019**, *11* (36), 33188–33193. https://doi.org/10.1021/acsami.9b07891.

(27) Guo, P.; Xu, J.; Gong, K.; Shen, X.; Lu, Y.; Qiu, Y.; Xu, J.; Zou, Z.; Wang, C.; Yan, H.; Luo, Y.; Pan, A.; Zhang, H.; Ho, J. C.; Yu, K. M. On-Nanowire Axial Heterojunction Design for High-Performance Photodetectors. *ACS Nano* **2016**, *10* (9), 8474–8481. https://doi.org/10.1021/acsnano.6b03458.

(28) Agarwal, R.; Lieber, C. M. Semiconductor Nanowires: Optics and Optoelectronics. *Appl. Phys. A* **2006**, *85* (3), 209. https://doi.org/10.1007/s00339-006-3720-z.

(29) Liu, X.; Long, Y.-Z.; Liao, L.; Duan, X.; Fan, Z. Large-Scale Integration of Semiconductor Nanowires for High-Performance Flexible Electronics. *ACS Nano* **2012**, *6* (3), 1888–1900. https://doi.org/10.1021/nn204848r.

(30) Tomioka, K.; Motohisa, J.; Hara, S.; Hiruma, K.; Fukui, T. GaAs/AlGaAs Core Multishell Nanowire-Based Light-Emitting Diodes on Si. *Nano Lett.* **2010**, *10* (5), 1639–1644. https://doi.org/10.1021/nl9041774.

(31) Winge, D. O.; Limpert, S.; Linke, H.; Borgström, M. T.; Webb, B.; Heinze, S.; Mikkelsen, A. Implementing an Insect Brain Computational Circuit Using III–V Nanowire Components in a Single Shared Waveguide Optical Network. *ACS Photonics* **2020**, *7* (10), 2787–2798. https://doi.org/10.1021/acsphotonics.0c01003.

(32) Fast, J.; Liu, Y.-P.; Chen, Y.; Samuelson, L.; Burke, A. M.; Linke, H.; Mikkelsen, A. Optical-Beam-Induced Current in InAs/InP Nanowires for Hot-Carrier Photovoltaics. *ACS Appl. Energy Mater.* **2022**, *5* (6), 7728–7734. https://doi.org/10.1021/acsaem.2c01208.

(33) Fast, J.; Aeberhard, U.; Bremner, S. P.; Linke, H. Hot-Carrier Optoelectronic Devices Based on Semiconductor Nanowires. *Applied Physics Reviews* **2021**, *8* (2), 021309. https://doi.org/10.1063/5.0038263.

(34) Wittenbecher, L.; Viñas Boström, E.; Vogelsang, J.; Lehman, S.; Dick, K. A.; Verdozzi, C.; Zigmantas, D.; Mikkelsen, A. Unraveling the Ultrafast Hot Electron Dynamics in Semiconductor Nanowires. *ACS Nano* **2021**, *15* (1), 1133–1144. https://doi.org/10.1021/acsnano.0c08101.

(35) Hajlaoui, C.; Pedesseau, L.; Raouafi, F.; Ben Cheikh Larbi, F.; Even, J.; Jancu, J.-M. First-Principles Calculations of Band Offsets and Polarization Effects at InAs/InP Interfaces. *J. Phys. D: Appl. Phys.* **2015**, *48* (35), 355105. https://doi.org/10.1088/0022-3727/48/35/355105.

(36) Faria Junior, P. E.; Campos, T.; Bastos, C. M. O.; Gmitra, M.; Fabian, J.; Sipahi, G. M. Realistic Multiband k · p Approach from *Ab Initio* and Spin-Orbit Coupling Effects of InAs and InP in Wurtzite Phase. *Phys. Rev. B* **2016**, *93* (23), 235204. https://doi.org/10.1103/PhysRevB.93.235204.

(37) Thelander, C.; Björk, M. T.; Larsson, M. W.; Hansen, A. E.; Wallenberg, L. R.; Samuelson, L. Electron Transport in InAs Nanowires and Heterostructure Nanowire Devices. *Solid State Communications* **2004**, *131* (9–10), 573–579. https://doi.org/10.1016/j.ssc.2004.05.033.

(38) Suyatin, D. B.; Thelander, C.; Björk, M. T.; Maximov, I.; Samuelson, L. Sulfur Passivation for Ohmic Contact Formation to InAs Nanowires. *Nanotechnology* **2007**, *18* (10), 105307. https://doi.org/10.1088/0957-4484/18/10/105307.

(39) Mead, C. A.; Spitzer, W. G. Fermi Level Position at Metal-Semiconductor Interfaces. *Phys. Rev.* **1964**, *134* (3A), A713–A716. https://doi.org/10.1103/PhysRev.134.A713.

(40) Bhargava, S.; Blank, H.-R.; Narayanamurti, V.; Kroemer, H. Fermi-Level Pinning Position at the Au–InAs Interface Determined Using Ballistic Electron Emission Microscopy. *Appl. Phys. Lett.* **1997**, *70* (6), 759–761. https://doi.org/10.1063/1.118271.

(41) Huang, Y.; Zhuge, F.; Hou, J.; Lv, L.; Luo, P.; Zhou, N.; Gan, L.; Zhai, T. Van Der Waals Coupled Organic Molecules with Monolayer MoS$_2$ for Fast Response Photodetectors with Gate-Tunable Responsivity. *ACS Nano* **2018**, *12* (4), 4062–4073. https://doi.org/10.1021/acsnano.8b02380.

(42) Politano, A.; Chiarello, G.; Samnakay, R.; Liu, G.; Gürbulak, B.; Duman, S.; Balandin, A. A.; Boukhvalov, D. W. The Influence of Chemical Reactivity of Surface Defects on Ambient-Stable InSe-Based Nanodevices. *Nanoscale* **2016**, *8* (16), 8474–8479. https://doi.org/10.1039/C6NR01262K.

(43) Riss, A.; Wickenburg, S.; Tan, L. Z.; Tsai, H.-Z.; Kim, Y.; Lu, J.; Bradley, A. J.; Ugeda, M. M.; Meaker, K. L.; Watanabe, K.; Taniguchi, T.; Zettl, A.; Fischer, F. R.; Louie, S. G.; Crommie, M. F. Imaging and Tuning Molecular Levels at the Surface of a Gated Graphene Device. *ACS Nano* **2014**, *8* (6), 5395–5401. https://doi.org/10.1021/nn501459v.

(44) Shu, H.; Cao, D.; Liang, P.; Jin, S.; Chen, X.; Lu, W. Effect of Molecular Passivation on the Doping of InAs Nanowires. *J. Phys. Chem. C* **2012**, *116* (33), 17928–17933. https://doi.org/10.1021/jp304350f.

(45) Du, J.; Liang, D.; Tang, H.; Gao, X. P. A. InAs Nanowire Transistors as Gas Sensor and the Response Mechanism. *Nano Lett.* **2009**, *9* (12), 4348–4351. https://doi.org/10.1021/nl902611f.

(46) Chen, Y.; Kivisaari, P.; Pistol, M.-E.; Anttu, N. Optimization of the Short-Circuit Current in an InP Nanowire Array Solar Cell through Opto-Electronic Modeling. *Nanotechnology* **2016**, *27* (43), 435404. https://doi.org/10.1088/0957-4484/27/43/435404.

(47) Yngman, S. GaN Nanowires as Probes for High Resolution Atomic Force and Scanning Tunneling Microscopy. *Rev. Sci. Instrum.* **2019**, 9.

(48) Schukfeh, M. I.; Storm, K.; Mahmoud, A.; Søndergaard, R. R.; Szwajca, A.; Hansen, A.; Hinze, P.; Weimann, T.; Svensson, S. F.; Bora, A.; Dick, K. A.; Thelander, C.; Krebs, F. C.; Lugli, P.; Samuelson, L.; Tornow, M. Conductance Enhancement of InAs/InP Heterostructure Nanowires by Surface Functionalization with

Oligo(Phenylene Vinylene)s. *ACS Nano* **2013**, *7* (5), 4111–4118. https://doi.org/10.1021/nn400380g.

(49) Yang, Y.; Peng, X.; Kim, H.-S.; Kim, T.; Jeon, S.; Kang, H. K.; Choi, W.; Song, J.; Doh, Y.-J.; Yu, D. Hot Carrier Trapping Induced Negative Photoconductance in InAs Nanowires toward Novel Nonvolatile Memory. *Nano Lett.* **2015**, *15* (9), 5875–5882. https://doi.org/10.1021/acs.nanolett.5b01962.

(50) Dayeh, S. A.; Soci, C.; Yu, P. K. L.; Yu, E. T.; Wang, D. Influence of Surface States on the Extraction of Transport Parameters from InAs Nanowire Field Effect Transistors. *Appl. Phys. Lett.* **2007**, *90* (16), 162112. https://doi.org/10.1063/1.2728762.

(51) Alexander-Webber, J. A.; Groschner, C. K.; Sagade, A. A.; Tainter, G.; Gonzalez-Zalba, M. F.; Di Pietro, R.; Wong-Leung, J.; Tan, H. H.; Jagadish, C.; Hofmann, S.; Joyce, H. J. Engineering the Photoresponse of InAs Nanowires. *ACS Appl. Mater. Interfaces* **2017**, *9* (50), 43993–44000. https://doi.org/10.1021/acsami.7b14415.

(52) Zhang, X.; Yao, X.; Li, Z.; Zhou, C.; Yuan, X.; Tang, Z.; Hu, W.; Gan, X.; Zou, J.; Chen, P.; Lu, W. Surface-States-Modulated High-Performance InAs Nanowire Phototransistor. *J. Phys. Chem. Lett.* **2020**, *11* (15), 6413–6419. https://doi.org/10.1021/acs.jpclett.0c01879.

(53) Han, Y.; Fu, M.; Tang, Z.; Zheng, X.; Ji, X.; Wang, X.; Lin, W.; Yang, T.; Chen, Q. Switching from Negative to Positive Photoconductivity toward Intrinsic Photoelectric Response in InAs Nanowire. *ACS Appl. Mater. Interfaces* **2017**, *9* (3), 2867–2874. https://doi.org/10.1021/acsami.6b13775.

(54) Pan, D.; Wang, J.-Y.; Zhang, W.; Zhu, L.; Su, X.; Fan, F.; Fu, Y.; Huang, S.; Wei, D.; Zhang, L.; Sui, M.; Yartsev, A.; Xu, H.; Zhao, J. Dimension Engineering of High-Quality InAs Nanostructures on a Wafer Scale. *Nano Lett.* **2019**, *19* (3), 1632–1642. https://doi.org/10.1021/acs.nanolett.8b04561.

# Supplementary Information for

# Combined Light Excitation and Scanning Gate Microscopy on Heterostructure Nanowire Photovoltaic Devices

Yen-Po Liu[1,3], Jonatan Fast[2,3], Yang Chen[2,3], Ren Zhe[1,3], Adam Burke[2,3], Rainer Timm[1,3], Heiner Linke[2,3], Anders Mikkelsen[1,3]

[1]*Division of Synchrotron radiation, Department of Physics, Lund University, Sweden*

[2]*Division of Solid State Physics, Department of Physics, Lund University, Sweden*

[3]*NanoLund, Lund University, Sweden*

## Table of contents

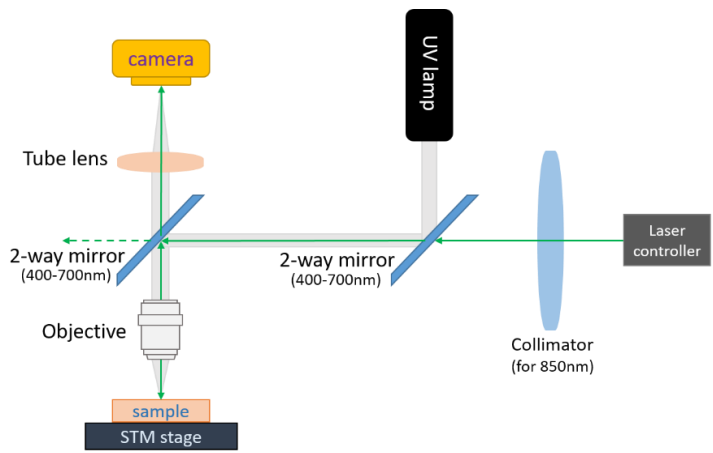

*Figure S1.* Optical parts of the SGM setup combined with laser excitation.

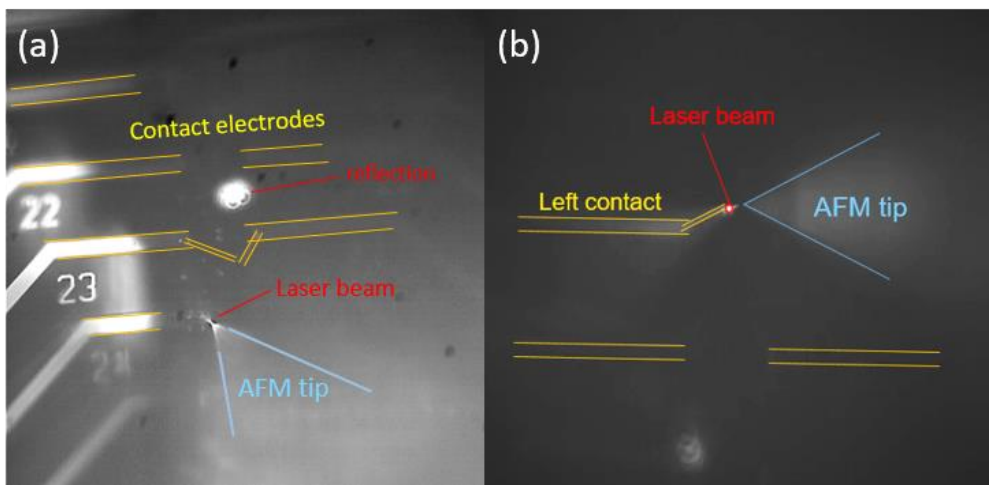

*Figure S2.* Camera view from the tilted optics showing the device contact electrodes, the laser beam and the AFM tip with different light conditions and beam positions. (a) A camera view when the beam is aligned on the tip, so left part of the electrodes are illuminated more. Also, a reflection spot is seen caused by 2 way mirrors and several lens of the optical system. (b) A zoom in view when the laser beam is focused on the device left electrode. A reduction of beam intensity is adjusted in order to see the tip and electrodes.

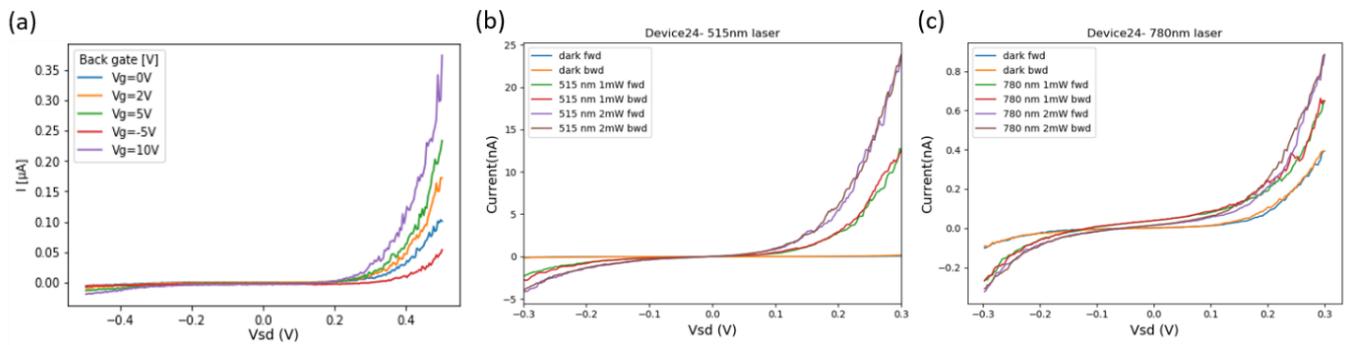

*Figure S3.* Electrical IVs of device II (a) IV sweep with back-gate (Si substrate) voltage -5 V, 0 V, 2 V, 5 V, and 10 V. (b) IV under dark and 515 nm laser of 1 mW and 2 mW. (c) IV under dark and 780 nm laser of 1 mW and 2 mW.

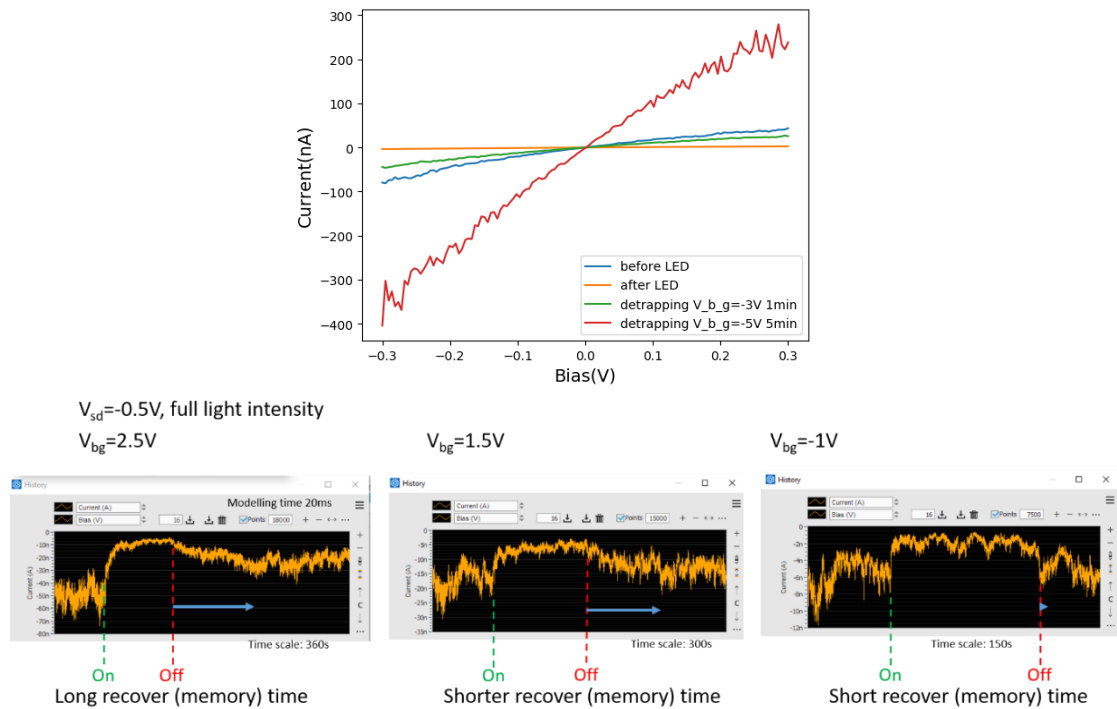

*Figure S4.* Trapping and de-trapping process on device I using the Si substrate as a backgate. TOP: I(V) curves of device before/after LED light illumination as well as after subsequent application of a back gate (detrapping). BOTTOM: Time traces of the effect of the light dependent on the applied backgate.

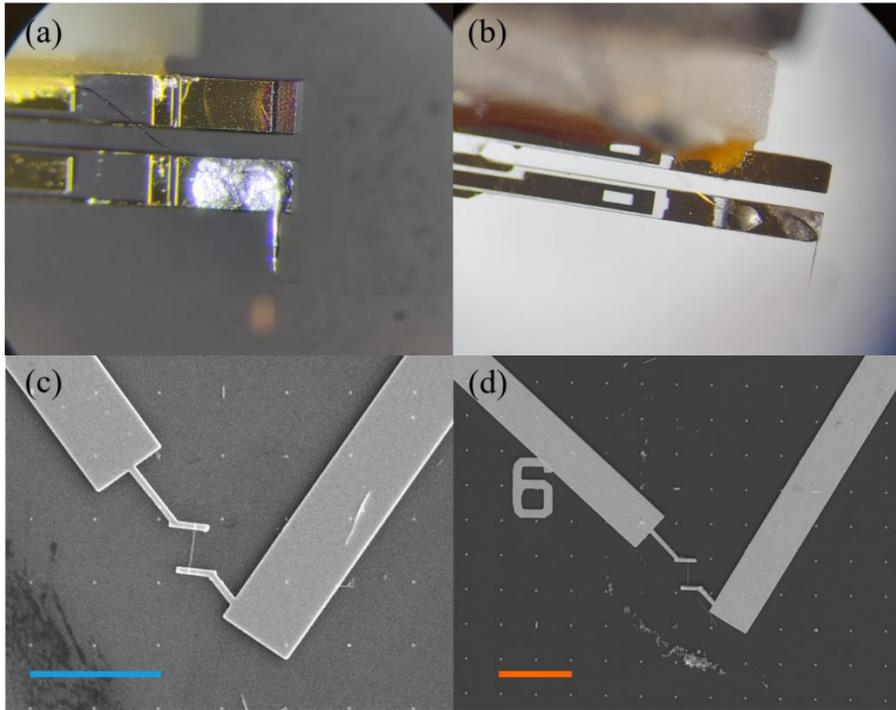

***Figure S5.*** *Images of the qPlus tips and InAs/InP NW devices for individual measurements (a) qPlus tip used for dark SGM measurement in Figure 2 and Figure 3. (b) qPlus tip used for laser combined SGM measurement in Figure 4 and Figure 5. (c) The InAs/InP NW device measured in the first part of the SGM dark data. (d) The InAs/InP NW device measured with laser combined SGM setup.*

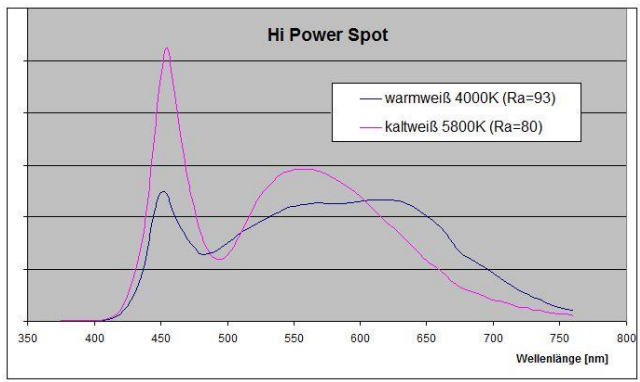

***Figure S6.*** *Spectrum of the Photonic Optics Hi power spot 5800K LED (light purple curve) used for LED lamp combined SGM measurement.*

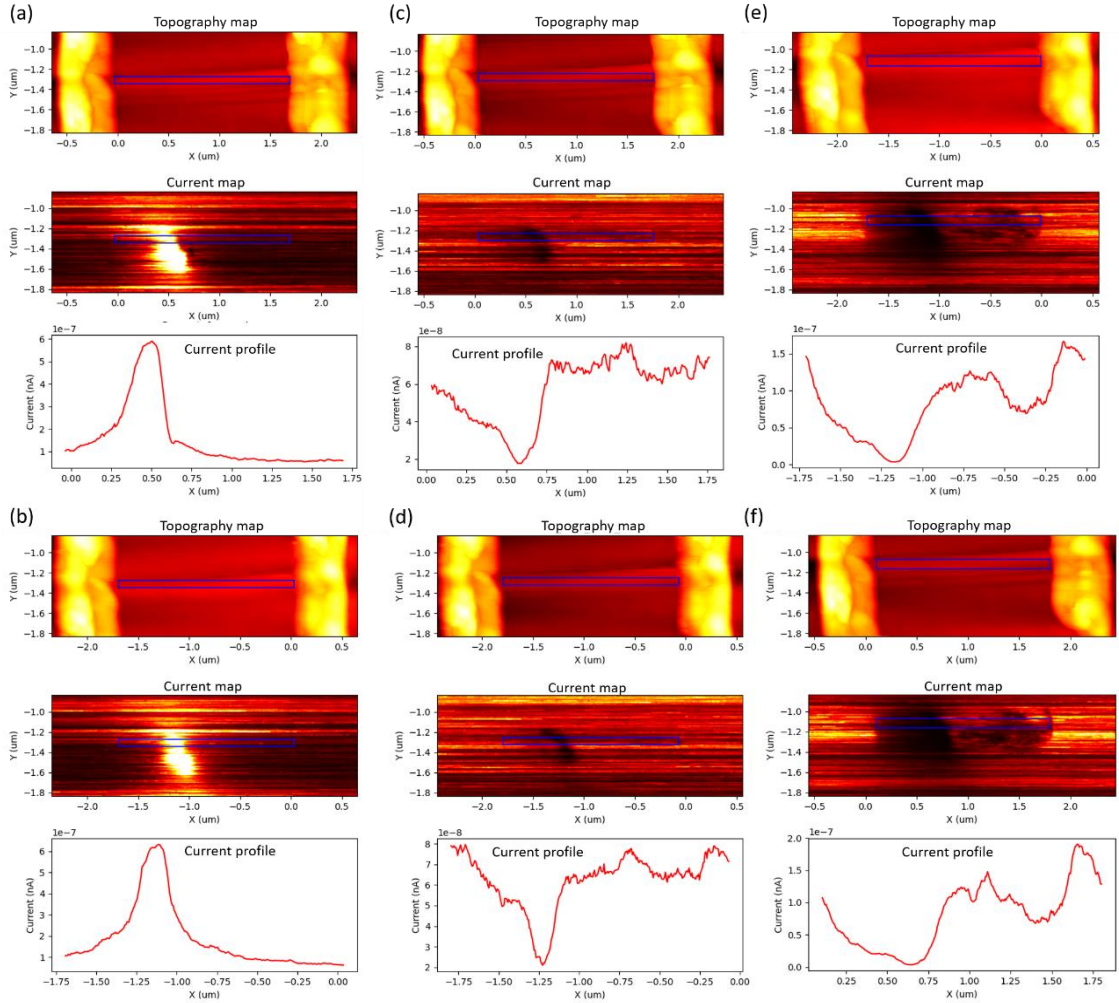

***Figure S7.*** *SGM data, showing Topography (top), current map (middle) and current profile of the selected area (bottom), measured with $V_{SD}=+0.6V$ in dark. (a) $V_{tip}=1.5\ V$ forward scan (b) $V_{tip}=1.5\ V$ backward scan (c) $V_{tip}=0\ V$ forward scan (d) $V_{tip}=0\ V$ backward scan (e) $V_{tip}=-1\ V$ forward scan (f) $V_{tip}=-1\ V$ backward scan.*

Here it can be noted that similar spectra were recorded with -0.6V bias across the wire yielding similar results (except at -1V) which is consisting with the observation that the wire barrier is symmetric already indicating that the overall observed behavior in this case originates from the barrier. While the device exposure to photons with energies much higher than the bandgap could pump the electrons from the valence band to a higher state and then fall into trapping states. The electrons sitting in the trapping state cannot migrate freely by the electrical potential gradient, contributing to the device transport. However, a negative tip bias, providing strong local electric field, would de-trap the electrons in the trapped states. Figure S4 shows the trapping and de-trapping processes observed during the measurement.